\documentclass[12pt]{iopart}
\usepackage{graphicx}
\usepackage{caption}
\usepackage{subcaption}

\begin{document}

\title[Measurement Back Action on Qubit Systems]{Measurement Back Action on Qubit Systems by a Cavity Probe}
\author{Yusui Chen$^{1,2}$, Wufu Shi$^2$, Debing Zeng$^2$, Ting Yu$^1$}

\address{$^1$Department of Physics and Center for Quantum Science and Engineering,\\
Stevens Institute of Technology, Castle Point on Hudson, Hoboken, NJ 07030 USA}

\address{$^2$Department of Applied Science and Techniques,\\
St. Peter's University, Jersey City, NJ 07306 USA}

\ead{ychen21@stevens.edu}
\vspace{10pt}
\begin{indented}
\item[]
\end{indented}

\begin{abstract}
We study the back action on a coupled multiple-qubit system induced by a quantum cavity probe in a non-demolition quantum measurement scheme. The exact quantum state stochastic Schr\"{o}dinger equation is derived to systematically investigate the dynamics of quantum entanglement of the multiple-qubit system inside the cavity probe. Although quantum entanglement cannot be directly measured through experiments, the back action on quantum entanglement in the multi-qubit system is witnessed by applying the exact quantum state stochastic equations. Our results demonstrate that the analysis on the sensitivity of the quantum measurement should include not only the standard quantum limit of the canonical operators, but also the back action on the quantum entanglement. Our new method proposed a theoretical approach to investigate the effects of the measurement back action on the multi-qubit systems induced by the cavity probe.
\end{abstract}

%
% Uncomment for keywords
%\vspace{2pc}
\noindent{\it Keywords}: Quantum back action, Quantum diffusion equations, Quantum measurement
%
% Uncomment for Submitted to journal title message
%\submitto{\JPA}
%
% Uncomment if a separate title page is required
%\maketitle
%
% For two-column output uncomment the next line and choose [10pt] rather than [12pt] in the \documentclass declaration
%\ioptwocol
%
\section{Introduction}
A substantial effort has been devoted to the development of quantum spin systems, used as the source of stationary qubits. Recently, a number of pioneering results have been proposed, spin read-out and preparation, as well as coherent spin control and spin-spin entanglement manipulation. In the study of solid-state quantum processing, both in theoretical and experimental, the central topic is how to effectively read out the quantum state of a spin system. Using a spin system coupled to a cavity, forming a cavity quantum electrodynamics (CQED) system, the efficiency of readout the spin system state can be enhanced significantly. Thus optical cavities potentially supply a platform to investigate fundamental concepts and quantum characters of quantum information processing, quantum open systems, quantum state preparation and quantum measurement \cite{cqed-bec,cqed-qd,cqed-block,cqed-block2,CQED1,CQED2,CQED3,CQED4,CQED5,CQED6}. A number of exciting results have been obtained, for instance, the deterministic multi-photon entangler inside a charged quantum dot micro-cavity \cite{cqed-1}, single shot initialization of a spin using a single optical pulse \cite{cqed-2}. Also the witness in experiments of the Bose-Einstein condensate, sing-photon blockade and n-photon blockade and so on. Finally, the spin-cavity interface can be applied to build up the quantum logic circuits and realize the quantum computation \cite{mqm1,mqm2}.

When take measurements on quantum systems, the non-demolition quantum measurement scheme is widely applied \cite{QM1}. As the accuracy of measurement increases, the objects behave quantum mechanically. One of the advantages of non-demolition quantum measurement is that this technique does not induce additional errors into the to-be-measured quantities and can achieve the standard quantum limit (SQL) in the measurement. It already has been successfully applied in many models \cite{GWD,GWD2,GWD3}. Since cavity QED can enhance the efficiency of detecting emitted photons and manipulating spin systems, we consider a non-demolition quantum measurement scheme on atomic systems by using the cavity probe, and the cavity is coupled to a photon detector. However the traditional discussion on such a non-demolition quantum measurement was previously developed for the weak coupling regime and isolated systems only, in which the cavity probe is treated as an idea quantum detector which behaves quantum mechanically during the measurement only and keeps classical and separated from the central spin system all the other time.

\begin{figure}
\centering
\includegraphics[width =.8 \linewidth]{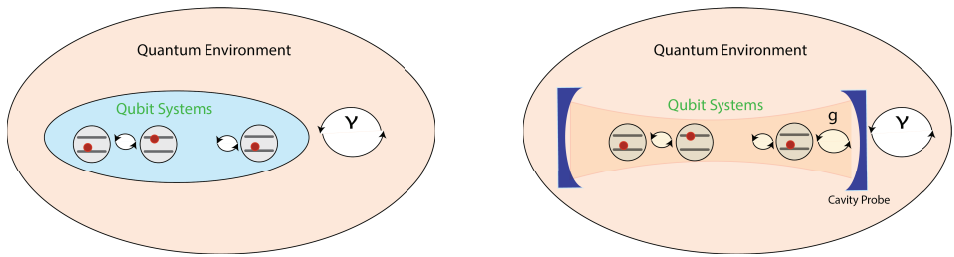}
\caption{Schematic of the multi-qubit system. Left graph shows the multi-qubit system is coupled to the environment directly. The right one represents a quantum measurement scheme on the qubit system via a quantum cavity probe.}
\label{model}
\end{figure}

We need to develop a theory to study the whole non-demolition quantum measurement scheme in the general framework of quantum optics and open quantum systems \cite{QO1,QO2,QO3,QO4}, as shown in Fig. \ref{model}. Thus the whole quantum system consists of three parts, the central spin system, the optical cavity probe and the photon detector. Under particular conditions, like the weak coupling strength case, the dynamics of the spin system can be approximately described by the Lindblad form master equation by taking the photon detector as a Markov environment. However, as we discussed, all the effort has been done is to increase the coupling strength between the spin system and the cavity probe, and enhance the ability to measure the spin systems. When the whole scheme is in the strong coupling regime, we need to identify respective contributions of the spin-spin coupling and of the quantum back action induced by the cavity probe on the dynamics of the spin system. Therefore, a more general theoretical tool is necessary to investigate the non-demolition quantum measurement scheme in the framework of non-Markovian open quantum systems.

In this paper, we apply the quantum-state diffusion (QSD) approach \cite{NM1,NM2,NM3,NM4} and extend it to the non-demolition quantum measurement scheme discussed above \cite{NM5,QM2,MKV1}. QSD approach, starting with the total system-environment Hamiltonian, supplies a clear microscopic explanation for the open quantum systems, and the derived QSD equations simulate the evolution of various quantum observables of the quantum system in the presence of the non-Markovian environment, in the dephasing and dissipative processing. One can generate the evolution of the reduced density matrix of the central spin system by taking ensemble average over all stochastic trajectories, which are governed by the QSD equations. Recently, it has been successfully applied in many models, such as multiple-qubit systems, multilevel atomic systems and coupled cavity systems \cite{NM6,NM7,chen1,chen2,NM8,chen3}. However, dealing with the hybrid system, spin system coupled to a cavity, in the open quantum systems is always a open question. For generality, we study the model without any approximation. However, the discussion on the back action and quantum measurement is limited only when the cavity probe and the photon detector are coupled in Markov regime. We demonstrate that the contribution of the cavity probe on the spin system can be revealed by taking multiple coupled stochastic processes. We provide comprehensive analytical derivation of the QSD equations of the spin system, in fully non-Markovian regime. In addition, our method can be extended to more complicated models, such as the spin system is embedded in a multi-layer cavities structure. We finally demonstrate that the modulation of quantum coherence and quantum entanglement due to the cavity probe, and the back action on the spin system induced by the cavity probe, which supplies an insight of the quantum computing, quantum coherence control and quantum information processing.

The paper is organized as follows. In section 2, we quickly review the standard QSD method and briefly
demonstrate the formal exact time-local non-Markovian QSD equations for the general non-demolition quantum measurement schemes. Our ansatz for this model is analytically derived. In section 3, we applied the exact QSD equations and the corresponding master equations to investigate the back action on the spin systems. Two examples are discussed, one-qubit and two-qubit systems. In the first example, we mainly focuses on the analytical derivation of the new method and its numerical simulation. In the second example, the influence of back-action on the quantum coherence and quantum entanglement in the spin system is demonstrated. In the last section, we come to some conclusions.

\section{Generate Model and Exact Quantum-State Diffusion (QSD) Equations}
\subsection{Review of standard QSD approach}
The total Hamiltonian of an open quantum system coupled to a zero-temperature bosonic environment can be described as, in the interaction picture, (setting $\hbar=1)$

\begin{equation}
H_{{\rm tot}}=H_{{\rm sys}}+L\sum_{k}g_{k}a_{k}^{\dagger}e^{i\omega_{k}t}+L^{\dagger}\sum_{k}g_{k}^{*}a_{k}e^{-i\omega_{k}t},
\end{equation}
where $L$ is the Lindblad operator of system and $a_{k}(a_{k}^{\dagger})$ is the annihilation (creation) operator of $k$th mode in the environment. Expanding the total quantum state in the Bargmann coherent state basis for all modes of the environment, $\langle z|=\langle z_{1},z_{2},...,z_{k},...|$, the $k$th mode of the environment is described by a coherent state with a random number $z_k$ and $a_{k}|z\rangle=z_{k}|z\rangle$. Thus the zero-temperature environment altogether can be described as a stochastic process, governed by the correlation function $\alpha(t,s)= \sum_k |g_k|^2 e^{i\omega_k (t-s)}$. Hence the stochastic Schr\"{o}dinger equation is shown as \cite{NM4}

\begin{equation}
\partial_{t}\psi_{t}(z^{*})=\left[-iH_{{\rm sys}}+Lz_{t}^{*}-L^{\dagger}\int_{0}^{t}{\rm d}s\alpha(t,s)\frac{{\rm \delta}}{\delta z_{s}^{*}}\right]\psi_{t}(z^{*}),\label{eq:original_qsd}
\end{equation}
where $\psi_{t}(z^{*})=\langle z|\psi_{{\rm tot}}\rangle$ is the
quantum trajectory.
$z_{t}^{*}=-i\sum_{k}g_{k}^{*}z_{k}^{*}e^{i\omega_{k}t}$ is the Gaussian
random process. For different frequency spectrum, $\alpha(t,s)$ takes the corresponding correlation function. The last term in above QSD equation,
functional derivative $\frac{\delta\psi_{t}}{\delta z_{s}^{*}}$,
can always rewritten in the form of a product of a time and noise dependent operator
$O(t,s,z^{*})$ and $\psi_{t}(z^{*})$, and $O$ operator can be
determined by the consistency condition, $\partial_{t}(\delta_{z_{s}^{*}}\psi_{t})=\delta_{z_{s}^{*}}(\partial_{t}\psi_{t}$).
Thus $O$ operator is determined by

\begin{equation}
\frac{\partial}{\partial t}O=[-iH_{{\rm sys}}+Lz_{t}^{*}-L^{\dagger}\bar{O},O]-L^{\dagger}\frac{\delta\bar{O}}{\delta z_{s}^{*}},\label{eq:O_consistency}
\end{equation}
and its initial condition $O(t,s=t)=L$, where $\bar{O}(t,z^{*})=\int_{0}^{t}ds\alpha(t,s)O(t,s,z^{*}).$ Once $O(t,s,z^{*})$
operator is obtained, the exact QSD equation can be written in a compact form
\begin{equation}
\partial_{t}\psi_{t}(z^{*})=\left[-iH_{{\rm sys}}+Lz_{t}^{*}-L^{\dagger}\bar{O}(t,z^{*})\right]\psi_{t}(z^{*}).
\end{equation}

\subsection{QSD equations for the spin system confined in a cavity probe}
The total Hamiltonian of the non-demolition quantum measurement scheme, consisting of three parts, the central quantum system $H_{\mbox{s }}$, the cavity
probe $H_{\mbox{p}}$ and the photon detector, as the environment $H_{\mbox{e}}$, is (setting $\hbar=1$)

\begin{equation}
H_{{\rm tot}} = H_{{\rm s}}+H_{{\rm sp}}+H_{{\rm p}}+H_{{\rm pe}}+H_{{\rm e}},\label{eq:total_H}
\end{equation}
\begin{equation}
H_{{\rm p}} =\sum_{k}\omega_{k}a_{k}^{\dagger}a_{k},\nonumber
\end{equation}
\begin{equation}
H_{{\rm e}}  =\sum_{k'}\omega_{k'}b_{k'}^{\dagger}b_{k'},\nonumber
\end{equation}
\begin{equation}
H_{{\rm sp}} =\sum_{k}(g_{k}La_{k}^{\dagger}+g_{k}^{*}L^{\dagger}a_{k}),\nonumber
\end{equation}
\begin{equation}
H_{{\rm pe}} =\sum_{j,k'}(f_{j,k'}L_{j}b_{k'}^{\dagger}+f_{j,k'}^{*}L^{\dagger}_{j}b_{k'}),\nonumber
\end{equation}
where $L$ is the Lindblad operator of the spin system. ${g_k}$ are the coupling strength between the spin system $k$th mode and the cavity probe. $L_{j}$ is the coupling operator between the cavity probe and the environment, and $f_{j,k'}$ is the corresponding coupling strength between the operator $L_{j}$ and the $k'$th mode in the environment. Based on the above Hamiltonian and the Markov approximation, the conventional Lindblad master equation can be obtained automatically \cite{QO1}. However for arbitrary coupling types, the Lindblad master equation is no longer valid. By contrary a theory considering the whole system in the framework of non-Markovian regime is necessary and the Markov case can be treated as a particular example.

Following the non-Markovian QSD approach, we treat the strongly coupled cavity and the non-Markovian environment as two
independent Gaussian noises, expanded in the Bargmann coherent state
basis $\langle z|a_{k}^{\dagger}=z_{k}^{*}\langle z|$ and $\langle y|b_{k'}^{\dagger}=y_{k'}^{*}\langle y|$. Since the cavity probe and the environment are two independent systems, therefore the two corresponding random variables $z_k^*$ and $y_{k'}^*$ are independent to each other. For simplicity, we consider a single mode cavity probe, linearly coupled to the environment. The formal QSD equation can be written as
\begin{eqnarray}
\partial_{t}\psi_{t} &=& \Biggl[-iH_{{\rm s}}+Lz_{t}^{*}-L^{\dagger}\int_{0}^{t}{\rm d}s\alpha(t,s)\frac{\delta}{\delta z_{s}^{*}}\Biggl]\psi_{t}\nonumber\\
& &+\Biggl[-iy_{t}^{*}\int_{0}^{t}{\rm d}s\alpha(t,s)\frac{\delta}{\delta z_{s}^{*}}-iz_{t}^{*}\int_{0}^{t}{\rm d}s\beta(t,s)\frac{\delta}{\delta y_{s}^{*}}\Biggl]\psi_{t},\label{eq:qsd_general}
\end{eqnarray}
where $\psi_{t}(z^{*},y^{*})=\langle z,y|\psi_{{\rm tot}}\rangle$ is the trajectory consisting of two independent noises.
Similarly, two Gaussian random processes $z_{t}^{*}$ and $y_{t}^{*}$ are defined as
\begin{eqnarray}
z_{t}^{*} =-i\sum_{k}g_{k}z_{k}^{*}e^{i\omega_{k}t},\ \
y_{t}^{*} =-i\sum_{k'}f_{k'}y_{k'}^{*}e^{i\omega_{k'}t},
\end{eqnarray}
with the corresponding correlation functions,
\begin{eqnarray}
\alpha(t,s) =\sum_{k}|g_{k}|^{2}e^{-i\omega_{k}(t-s)},\ \
\beta(t,s) =\sum_{k'}|f_{k'}|^{2}e^{-i\omega_{k'}(t-s)}.
\end{eqnarray}
Note that the correlation functions in our discussion are directly determined by the frequency spectrum. For each stochastic process, we define the corresponding $O$ operator
\begin{eqnarray}
\frac{\delta\psi_{t}}{\delta z_{s}^{*}} =O_{z}(t,s,z^{*},y^{*})\psi_{t},\ \
\frac{\delta\psi_{t}}{\delta y_{s}^{*}} =O_{y}(t,s,z^{*},y^{*})\psi_{t}.
\end{eqnarray}
Both $O_{z}$ and $O_{y}$ operators consist of the two noises $z^{*}$ and $y^{*}$ at the same time. The time evolution is
determined by the consistency conditions, $\partial_{t}(\delta_{z_{s}^{*}}\psi_{t})=\delta_{z_{s}^{*}}(\partial_{t}\psi_{t})$
and $\partial_{t}(\delta_{y_{s}^{*}}\psi_{t})=\delta_{y_{s}^{*}}(\partial_{t}\psi_{t}$) respectively. So that, we have
\begin{equation}
\partial_{t}O_{z} =\left[-iH_{eff},\,O_{z}\right]-L^{\dagger}\frac{\delta\bar{O}_{z}}{\delta z_{s}^{*}}-iy_{t}^{*}\frac{\delta\bar{O}_{z}}{\delta z_{s}^{*}}-iz_{t}^{*}\frac{\delta\bar{O}_{y}}{\delta z_{s}^{*}},\label{eq:consistency}\nonumber
\end{equation}
\begin{equation}
\partial_{t}O_{y} =\left[-iH_{eff},\,O_{y}\right]-L^{\dagger}\frac{\delta\bar{O}_{z}}{\delta y_{s}^{*}}-iy_{t}^{*}\frac{\delta\bar{O}_{z}}{\delta y_{s}^{*}}-iz_{t}^{*}\frac{\delta\bar{O}_{y}}{\delta y_{s}^{*}},\nonumber
\end{equation}
where
\begin{equation}
H_{eff}=H_{{\rm s}}+iLz_{t}^{*}-iL^{\dagger}\bar{O}_{z}+y_{t}^{*}\bar{O}_{z}+z_{t}^{*}\bar{O}_{y}.
\end{equation}
$\bar{O}_{z}$ and $\bar{O}_{y}$ are defined as $\bar{O}_{z}=\int_{0}^{t}ds\alpha(t,s)O_{z}$ and $\bar{O}_{y}=\int_{0}^{t}ds\beta(t,s)O_{y}$.
When cut off the outer layer environment, the above derivation can be downgraded to the standard QSD equation. Since $y_{t}^{*}$ is eliminated
and the corresponding correlation function $\beta(t,s)$ is zero, and the above effective Hamiltonian is degenerated into $H_{eff} =H_{{\rm s}}+iLz_{t}^{*}-iL^{\dagger}\bar{O}$, the standard form in QSD approach. Although the two stochastic processes are independent, the two $O$ operators are correlated as the cavity probe and the environment are coupled. The two $O$ operators are determined by
\begin{equation}
O_{z}(t,s,z^{*}) =O(t,s)-i\int_{0}^{s}{\rm d}\tau\beta(t,\tau)O_{y}(t,\tau,z^{*}),\nonumber
\end{equation}
\begin{equation}
O_{y}(t,s,z^{*}) =-i\int_{0}^{s}{\rm d}\tau\alpha(t,\tau)O_{z}(t,\tau,z^{*}),\label{eq:new_ansatz}
\end{equation}
where $O(t,s)$ is the conventional $O$ operator when the
spin system is coupled to the environment directly. In order to satisfy
the consistency conditions (\ref{eq:consistency}), $O_{z}$
and $O_{y}$ must consist of the same operator basis as $O(t,s)$.
Notably, the up limit of the integration in $O_{z}$ and $O_{y}$ is time index $s$, not $t$.
This definition naturally matches with the physics in the process that the non-Markovian effort consists of two  parts: 1) the coupling between the spin system and the cavity, the first layer; 2) the coupling between the cavity and the surrounding environment, the second layer. These two couplings supply two channels to allow the energy flow in and out. In addition, we can check the Markov limit by setting the two correlation functions as Dirac delta function, then $O_{y}=0$ and $O_{z}=L$, which matches the conclusion of Markov QSD equations. Next, we will take some examples and demonstrate analytical derivation.

\section{Examples and Numerical Results}
\subsection{Back action on one qubit system}
As mentioned in the Fig. 1, we take one qubit system as our first
example. For simplicity, we use the one mode cavity as the quantum
probe. The total Hamiltonian (\ref{eq:total_H}) is written as (setting $\hbar =1$)
\begin{equation}
H_{{\rm tot}}=H_{{\rm qc}}+H_{\rm{int}}+H_{{\rm env}},
\end{equation}
\begin{equation}
H_{{\rm qc}} =\frac{\omega_{s}}{2}\sigma_{z}+gLa^{\dagger}+g^{*}L^{\dagger}a+\omega a^{\dagger}a,\nonumber
\end{equation}
\begin{equation}
H_{{\rm env}} =\sum_{k}\omega_{k}b_{k}^{\dagger}b_{k,}\nonumber
\end{equation}
\begin{equation}
H_{{\rm int}} =\sum_{k}(f{}_{k}ab_{k}^{\dagger}+f_{k}^{*}a^{\dagger}b_{k}),
\end{equation}
where $L=\sigma_{-}$ is the system Lindblad operator. $g$ is the
coupling strength between the qubit and cavity, and $f_{k}$ are coupling
constants between cavity and environment. As we know, in such simple
example, the $O$ operator is follow the form of $f(t,s)\sigma_{-}$
\cite{NM4} and the QSD equation can be written in the form of
Eq. (\ref{eq:qsd_general})
\[
\partial_{t}\psi_{t}=\left[-i\frac{\omega_{s}}{2}\sigma_{z}+Lz_{t}^{*}-(L^{\dagger}+iy_{t}^{*})\bar{O}_{z}-iz_{t}^{*}\bar{O}_{y}\right]\psi_{t},
\]
where $z_{t}^{*}=-igz^{*}e^{i\omega t}$ is the first complex Gaussian
noise and its correlation function $\alpha(t,s)=|g|^{2}e^{-i\omega(t-s)}=\mathcal{M}[z_{t}z_{s}^{*}],$
for the initial state of the cavity is prepared as a vacuum state.
The second noise $y_{t}^{*}=-i\sum_{k}\frac{f{}_{k}}{g^{*}}y_{k}^{*}e^{i\omega_{k}t}$
and its correlation function is $\beta(t,s)=\frac{1}{\mathbf{|g|^{2}}}\sum_{k}|f_{k}|^{2}e^{-i\omega_{k}(t-s)}=\mathcal{M}[y_{t}y_{s}^{*}]$. The symbol $\mathcal{M}[\cdot]=\int\frac{d^{2}z}{\pi}\frac{d^{2}y}{\pi}e^{-|z|^{2}}e^{-|y|^{2}}\cdot$
is the ensemble average over two noises. In addition, $\bar{O}_{z}=\int_{0}^{t}{\rm d}s\alpha(t,s)O_{z}(t,s)$
and $\bar{O}_{y}=\int_{0}^{t}{\rm d}s\beta(t,s)O_{y}(t,s)$. Assuming
the coefficient functions are $n(t,s)$ and $m(t,s)$ respectively,
\begin{equation}
O_{z}=n(t,s)\sigma_{-} =f(t,s)\sigma_{-}-i\int_{0}^{s}{\rm d}\tau\beta(t,\tau)m(t,\tau)\sigma_{-},\nonumber
\end{equation}
\begin{equation}
O_{y}=m(t,s)\sigma_{-} =-i\int_{0}^{s}{\rm d}\tau\alpha(t,\tau)n(t,\tau)\sigma_{-}.\label{eq:ansatz_JC}
\end{equation}
By definition, the corresponding coefficient functions $N(t)$ and
$M(t)$ in $\bar{O}_{z}$ and $\bar{O}_{y}$ operators,
\begin{equation}
N(t) =\int_{0}^{t}{\rm d}s\alpha(t,s)n(t,s),
\end{equation}
\begin{equation}
M(t) =\int_{0}^{t}{\rm d}s\beta(t,s)m(t,s).
\end{equation}
Substitute the new ansatz Eq. \ref{eq:ansatz_JC} into $O$ operator
consistency conditions, we obtain a system of two partial differential
equations
\begin{equation}
\frac{\partial n}{\partial t} =i\omega_{s}n+Nn,
\end{equation}
\begin{equation}
\frac{\partial m}{\partial t} =i\omega_{s}m+Nm.\label{eq:coeff_jc}
\end{equation}
If we take the time boundary condition of $s\rightarrow t$, we can generate the new initial conditions as,
\begin{equation}
n(t,t) =1-iM(t),
\end{equation}
\begin{equation}
m(t,t) =-iN(t),\label{eq:initial_jc}
\end{equation}
where we use the initial value $f(t,t)=1$. We can solve $N(t)$ and
$M(t)$ numerically for arbitrary correlation function by Eq. \ref{eq:coeff_jc}
with the boundary condition Eq. \ref{eq:initial_jc}. For simplicity,
we choose Ornstein-Uhlenbeck environmental noise, with $\beta(t,s)=\frac{\gamma}{2|g|^{2}}e^{-\gamma|t-s|}$.
When $\gamma\rightarrow0$, the correlation function has very long
memory time, equivalent to strong non-Markovian case. By contrary
if $\gamma\rightarrow\infty$, $\beta(t,s)\rightarrow\delta(t,s)$
approaches to the Markov limit.
\begin{equation}
\frac{{\rm d}N}{{\rm d}t} =|g|^{2}(1-iM)+i(\omega_{s}-\omega)N+N^{2},
\end{equation}
\begin{equation}
\frac{{\rm d}M}{{\rm d}t} =\frac{-i\gamma}{2|g|^{2}}N+(i\omega_{s}-\gamma)M+NM.
\end{equation}
With the explicit $O$ operator \ref{eq:ansatz_JC}, the exact
time local QSD equation can be written in an appealing form,
\begin{equation}
\partial_{t}\psi_{t}=\left[-i\frac{\omega_{s}}{2}\sigma_{z}+\sigma_{-}z_{t}^{*}-N\sigma_{+}\sigma_{-}-i(Ny_{t}^{*}+Mz_{t}^{*})\sigma_{-}\right]\psi_{t}.
\end{equation}
Quantum state trajectory recovers density matrix after ensemble average:
$\rho_{t}=\mathcal{M}[|\psi_{t}\rangle\langle\psi_{t}|]$. By Novikov
theorem \cite{NM8},
\begin{equation}
\mathcal{M}[z_{t}^{*}|\psi_{t}\rangle\langle\psi_{t}|] =\rho_{t}\bar{O}_{z}^{\dagger},\\
\mathcal{M}[y_{t}^{*}|\psi_{t}\rangle\langle\psi_{t}|] =\rho_{t}\bar{O}_{y}^{\dagger},
\end{equation}
Then we can derive the exact master equation for this model for this simple case. General derivation of non-Markovian master equations for multiple-qubit systems can follow the ref. \cite{Chen1}.
\begin{equation}
\partial_{t}\rho_{t}=-i[\frac{\omega_{s}}{2}\sigma_{z},\rho_{t}]+[\sigma_{-},\rho_{t}\bar{O}_{z}^{\dagger}]-[\sigma_{+},\bar{O}_{z}\rho_{t}].
\end{equation}

\begin{figure}
\centering
\includegraphics[width =.8 \linewidth]{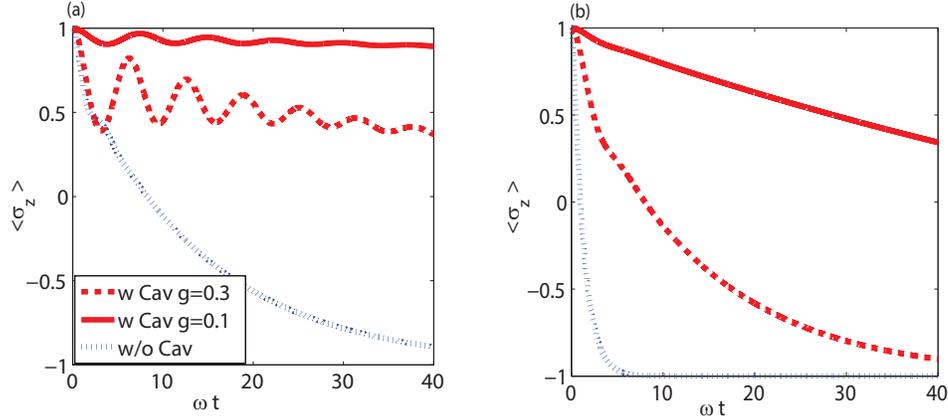}
\caption{Dynamics of decay for the qubit from initial state $\psi_{0}=|1\rangle$.
Initial state of the cavity is set as vacuum state. (a) shows
non-Markovian regime with $\gamma=0.5$, and (b) shows
Markovian case with $\gamma=5$. Red curves (solid and dashed) show
single qubit coupled to cavity probe, indirect measurement. Blue (dotted)
one shows single qubit coupled to environment without cavity probe,
direct measurement, for comparison. The parameters are: $\omega_{s}=2\omega_{cav}=\omega$.}
\label{JC}
\end{figure}

In Fig.\ref{JC}, the dynamics of the average population of the qubit system embedded in the cavity (red curves) is plotted. Because it is not easy to quantitatively investigate the back action on the atomic on system due to the cavity probe, we simply show the dynamics of atomic systems without the cavity (blue curves), as a comparison. In Fig. \ref{JC} (a), the dynamics is generically discussed when the system is in the presence of a non-Markovian environment. The influence of the cavity performs as the manipulation of the coherence time. In Fig. \ref{JC} (b), the decaying factor $\gamma=5$, and is in the Markov regime, equivalent to a real quantum measurement scheme. Although all evolution curves show the Markov-like decaying, the back action on the qubit system is witnessed, longer decaying time and lower energy-loss rate. For the coupling between qubit and cavity $g$ (red solid line and red dashed line), higher $g$-value creates the high-efficiency energy transition channel which allows the energy quanta in qubit transit in and out the cavity faster. This study informed us that the back action on coherence time is a must considered element when we perform the time-dependent measurements.

In this simple example, we focus on how to derive the new non-Markovian
QSD equations for the embedded atomic system. Principally, the $O$ operator
in regular QSD approach must be used, since the new $O_{z}$ and $O_{y}$
operators can be linearly decomposed into an operator basis, same as
the regular $O$ operator. Also we showed the possibility of deriving
master equations from corresponding QSD equations by applying Novikov
theorem.

\subsection{Back action on two-qubit systems}
One qubit case is simple in math and physical phenomenon, as the decoherence evolution. In this section, we study the two-qubit system, on which the back action is investigated. In the two-qubit system, more non-linear quantum characters will be explored. The total Hamiltonian is shown as,
\begin{equation}
H_{{\rm tot}}=H_{{\rm q}}+H_{{\rm qc}}+H_{{\rm c}}+H_{{\rm int}}+H{\rm _{{\rm env}}},
\end{equation}
\begin{equation}
H_{{\rm q}} =\frac{\omega_{s}}{2}\left(\sigma_{z}^{A}+\sigma_{z}^{B}\right),
\end{equation}
\begin{equation}
H_{{\rm c}} =\omega a^{\dagger}a,
\end{equation}
\begin{equation}
H_{{\rm qc}} =gLa^{\dagger}+g^{*}L^{\dagger}a,
\end{equation}
\begin{equation}
H_{{\rm int}} =a\sum_{k}f_{k}b_{k}^{\dagger}+a^{\dagger}\sum_{k}f_{k}^{*}b_{k},
\end{equation}
\begin{equation}
H_{\rm{env}} =\sum_{k}\omega_{k}b_{k}^{\dagger}b_{k},
\end{equation}
where $L=\kappa_{1}\sigma_{-}^{A}+\kappa_{2}\sigma_{-}^{B}$ is the
system Lindblad operator. Recent research has shown that the $O$
operator for two-qubit dissipative model always can be decomposed
into five operators as basis \cite{chen1}:
\begin{equation}
O =\sum_{j=1}^{4}f_{j}(t,s)O_{j}+i\int_{0}^{t}{\rm d}s'f_{5}(t,s,s')z_{s'}^{*}O_{5},\label{eq:2spin_original_O}
\end{equation}
\begin{equation}
O_{1} =\sigma_{-}^{A},\,O_{2}=\sigma_{-}^{B},
\end{equation}
\begin{equation}
O_{3}=\sigma_{z}^{A}\sigma_{-}^{B},
O_{4}=\sigma_{-}^{A}\sigma_{z}^{B},\,O_{5}=\sigma_{-}^{A}\sigma_{-}^{B}.\nonumber
\end{equation}
By substituting (\ref{eq:2spin_original_O}) into consistency condition,
all coefficients could be determined, with its boundary condition, as shown in the appendix. Note that this exact $O$ operator consists of noise term, so that $O_{z}$ and $O_{y}$ can be expanded by the regular operator basis, consisting of two noises:
\begin{eqnarray}
O_{z} =\sum_{j=1}^{4}n_{j}(t,s)O_{j}
  +i\int_{0}^{t}{\rm d}s'\left(n_{5}(t,s,s')z_{s'}^{*}+n_{6}(t,s,s')y_{s}^{*}\right)O_{5}, \nonumber \\
O_{y} =\sum_{j=1}^{4}m_{j}(t,s)O_{j}
  +i\int_{0}^{t}{\rm d}s'\left(m_{5}(t,s,s')z_{s'}^{*}+m_{6}(t,s,s')y_{s}^{*}\right)O_{5} \label{eq:2spin_O}
\end{eqnarray}
Above coefficient functions could be numerically evaluated by a group of differential equations, and we
can determine the exact $O$ operator (\ref{eq:2spin_O}) thereafter. Finally
we write down the time-local QSD equation in the formalism of $O_z$ and $O_y$:
\begin{equation}
\partial_{t}\psi_{t}=\left[-iH_{{\rm s}}+Lz_{t}^{*}-\left(L^{\dagger}+iy_{t}^{*}\right)\bar{O}_{z}-iz_{t}^{*}\bar{O}_{y}\right]\psi_{t}
\end{equation}

\begin{figure} [htbp]
\centering
\includegraphics[width=.8 \linewidth]{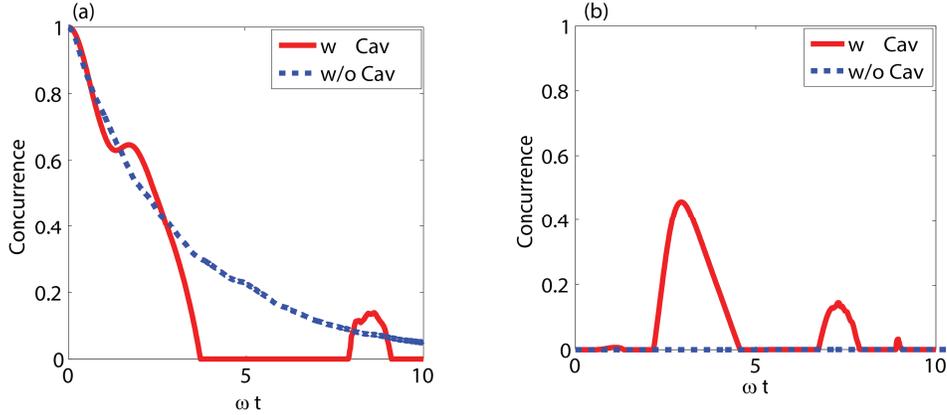} %100 percent
\caption{Dynamics of entanglement of two qubits. Indirect measurement (red
solid) and direct measurement (blue dashed) cases are drawn separately.
In (a), the qubit system initial state is $(|11\rangle+|00\rangle)/\sqrt{2}$
and the initial state of the cavity probe is vacuum state. In (b),
the qubit system initial state is $|11\rangle$. The parameters are
set as $\omega_{s}=2\omega_{cav}=\omega$, $g=0.5\omega$, $\gamma=5,$
$\kappa_{1}=\kappa_{2}=1$.}
\label{2spin}
\end{figure}

In precise quantum measurement schemes, the discussions on sensitivity usually are focused on the standard quantum limit (SQL) of particular canonical quantum operators and their Heisenberg uncertainty limit. However, the error analysis on some quantities like quantum entanglement is invalid, because quantum entanglement is not an operator, and cannot be directly observed in the experiment. Any experimental analysis on quantum entanglement is rooted on the reduced density matrix gained via quantum tomography. In order to theoretically investigate the back action on such non-linear quantities, we simulate their evolution (red solid line) by using the extended QSD approach. In addition, we select a control group (blue dashed line) in which the atomic system is coupled directly to the detector. We solve the generic non-Markovian evolution, but we limit the discussion on back action in Markov regime, by taking the decay coefficient $\gamma=5$. In Fig. \ref{2spin} (a), we prepare the qubit system in the maximally entangled Bell state $(|11\rangle+|00\rangle)/\sqrt{2}$. Entanglement fluctuation and entanglement sudden death \cite{NM1}
and rebirth are observed. In contrary, the qubit system comes to disentangled eventually without rebirth when it is coupled to the detector directly.

In Fig. \ref{2spin} (b), the two-qubit system is prepared in a separate state $|11\rangle$ initially, and the coefficients are set as $\kappa_{1}=\kappa_{2}=1$. For this initial state, the two-qubit system should keep separated if it is directly coupled to the detector, and this phenomenon is shown (blue dashed line). However, we also notice that quantum entanglement arises when the cavity probe involves in the measurement. In another word, the generated quantum entanglement is the back action on the two-qubit system purely induced by the cavity probe. This result inspires us that even in a Markov measurement scheme, the system evolution still displays non-Markovian characters and long memory-time effect. Additionally, once we perform a sequential measurement on a quantum system, the deflection from the internal entanglement should be considered. Because of the strong coupling between the qubit system and the cavity probe, the collective evolution of the atomic system still show the non-Markovian characters.

\section{Conclusion}

In conclusion, the purpose of this paper is to analytically investigate
the back action on the atomic systems in a non-demolition quantum measurement scheme
via a cavity probe, particularly the back action of the nonlinear quantities, for example, quantum coherence and quantum entanglement. To solve this model analytically, we extend the regular QSD approach and derive the exact QSD equation in the space of two independent noises. The technical difficulties come from that the interactions between the quantum probe and the environment are in the strong regime, therefore the approximation methods based on weak coupling assumptions are no longer valid. We then derive a new set of $O$ operators consisting of two independent noises respectively and determine the corresponding non-Markovian QSD equations, which allow us to investigate the quantum entanglement dynamics via the numerical simulations. In the two-qubit example, we study the back action on the quantum entanglement due to the cavity quantum probe, which inspires us that precision analysis is not complete if we only discuss the SQL of the canonical operators. The back action induced by the quantum probe on the undetectable quantities is also of great importance and necessity. Especially, to some non-observable quantities, our theoretical method supplies a possibility to study the back action and its influences.

\appendix
\section{Coefficients for QSD equations}
In order to determine the coefficients in the $O$ operators, we simply substitute Eq. (\ref{eq:2spin_O}) into Eq. (\ref{eq:consistency}), and
obtain a group of partial differential equations for all coefficients.
\begin{eqnarray}
\partial_{t}n_{1} &=&i\omega_{A}n_{1}+\kappa_{1}N_{1}n_{1}+\kappa_{1}N_{4}n_{4}-\kappa_{2}N_{1}n_{3}+\kappa_{2}N_{3}n_{1}+\kappa_{2}N_{3}n_{4}+\kappa_{2}N_{4}n_{3} \nonumber\\
& &-\frac{iN_{5}}{2}\kappa_{2}, \\
\partial_{t}n_{2} &=&i\omega_{B}n_{2}-\kappa_{1}N_{2}n_{4}+\kappa_{1}N_{3}n_{4}+\kappa_{1}N_{4}n_{2}+\kappa_{1}N_{4}n_{3}+\kappa_{2}N_{2}n_{2}+\kappa_{2}N_{3}n_{3}\nonumber\\
& &-\frac{iN_{5}}{2}\kappa_{1},\\
\partial_{t}n_{3} &=&i\frac{\omega_{B}}{2}n_{3}-\kappa_{1}N_{2}n_{1}+\kappa_{1}N_{3}n_{1}+\kappa_{1}N_{4}n_{2}+\kappa_{1}N_{4}n_{3}+\kappa_{2}N_{2}n_{3}+\kappa_{2}N_{3}n_{2}\nonumber\\
& &-\frac{iN_{5}}{2}\kappa_{1},\\
\partial_{t}n_{4} &=&i\frac{\omega_{A}}{2}n_{4}+\kappa_{1}N_{1}n_{4}+\kappa_{1}N_{4}n_{1}-\kappa_{2}N_{1}n_{2}+\kappa_{2}N_{3}n_{1}+\kappa_{2}N_{3}n_{4}+\kappa_{2}N_{4}n_{2}\nonumber\\
& &-\frac{iN_{5}}{2}\kappa_{2},\\
\partial_{t}n_{5} &=&i(\omega_{A}+\omega_{B})n_{5}+\kappa_{1}N_{1}n_{5}+\kappa_{1}N_{4}n_{5}+\kappa_{1}N_{5}\left(n_{1}-n_{4}\right)\nonumber\\
& &+\kappa_{2}N_{2}n_{5}+\kappa_{2}N_{3}n_{5}+\kappa_{2}N_{5}(n_{2}-n_{3}),
\\
\partial_{t}n_{6} &=&i(\omega_{A}+\omega_{B})n_{6}+\kappa_{1}N_{1}n_{6}+\kappa_{1}N_{4}n_{6}+\kappa_{1}N_{6}\left(n_{1}-n_{4}\right)\nonumber\\
& &+\kappa_{2}N_{2}n_{6}+\kappa_{2}N_{3}n_{6}+\kappa_{2}N_{6}(n_{2}-n_{3}),
\\
\partial_{t}m{}_{1} &=&i\omega_{A}m_{1}+\kappa_{1}N_{1}m_{1}+\kappa_{1}N_{4}m_{4}-\kappa_{2}N_{1}m_{3}+\kappa_{2}N_{3}m_{1}+\kappa_{2}N_{3}m_{4}+\kappa_{2}N_{4}m_{3}\nonumber\\
& &-\frac{iN_{6}}{2}\kappa_{2},
\\
\partial_{t}m_{2} &=&i\omega_{B}m_{2}-\kappa_{1}N_{2}m_{4}+\kappa_{1}N_{3}m_{4}+\kappa_{1}N_{4}m_{2}+\kappa_{1}N_{4}m_{3}+\kappa_{2}N_{2}m_{2}+\kappa_{2}N_{3}m_{3}\nonumber\\
& &-\frac{iN_{6}}{2}\kappa_{1},
\\
\partial_{t}m_{3} &=&i\frac{\omega_{B}}{2}m_{3}-\kappa_{1}N_{2}m_{1}+\kappa_{1}N_{3}m_{1}+\kappa_{1}N_{4}m_{2}+\kappa_{1}N_{4}m_{3}+\kappa_{2}N_{2}m_{3}+\kappa_{2}N_{3}m_{2}\nonumber\\
& &-\frac{iN_{6}}{2}\kappa_{1},
\\
\partial_{t}m_{4} &=&i\frac{\omega_{A}}{2}m_{4}+\kappa_{1}N_{1}m_{4}+\kappa_{1}N_{4}m_{1}-\kappa_{2}N_{1}m_{2}+\kappa_{2}N_{3}m_{1}+\kappa_{2}N_{3}m_{4}+\kappa_{2}N_{4}m_{2}\nonumber\\
& &-\frac{iN_{6}}{2}\kappa_{2},
\\
\partial_{t}m_{5} &=&i(\omega_{A}+\omega_{B})m_{5}+\kappa_{1}N_{1}m_{5}+\kappa_{1}N_{4}m_{5}+\kappa_{1}N_{5}\left(m_{1}-m_{4}\right)\nonumber\\
& &+\kappa_{2}N_{2}m_{5}+\kappa_{2}N_{3}m_{5}+\kappa_{2}N_{5}(m_{2}-m_{3}),
\\
\partial_{t}m_{6} &=&i(\omega_{A}+\omega_{B})m_{6}+\kappa_{1}N_{1}m_{6}+\kappa_{1}N_{4}m_{6}+\kappa_{1}N_{6}\left(m_{1}-m_{4}\right)\nonumber\\
& & +\kappa_{2}N_{2}m_{6}+\kappa_{2}N_{3}m_{6}+\kappa_{2}N_{6}(m_{2}-m_{3}),
\end{eqnarray}
where $M_{j}(t)=\int_{0}^{t}{\rm d}s\beta(t,s)m_{j}(t,s)$ and $N_{j}(t)=\int_{0}^{t}{\rm d}s\alpha(t,s)n_{j}(t,s)\,\,(j=1,2,3,4)$. Similarly, we can define $N_5(t,s'), N_6(t,s'), M_5(t,s'), M_6(t,s')$ as
\begin{eqnarray}
N_{5,6}(t,s')=i\int_{0}^{t}{\rm d}s'\alpha(t,s)n_{5,6}(t,s,s'),\\
M_{5,6}(t,s')=i\int_{0}^{t}{\rm d}s'\beta(t,s)m_{5,6}(t,s,s').
\end{eqnarray}
Meanwhile we have the boundary conditions:
\begin{eqnarray}
n_{5}(t,s,t) &=& -2i(\kappa_{1}n_{3}+\kappa_{2}n_{4})-iM_{5} \nonumber \\
 & &-2(M_{1}n_{3}+M_{2}n_{4}-M_{3}n_{1}-M_{4}n_{2}),
\\
n_{6}(t,s,t) &=&-iN_{5},
\\
m_{5}(t,s,t) &=&-2i(\kappa_{1}m_{3}+\kappa_{2}m_{4})-iM_{6},
\\
m_{6}(t,s,t) &=&-iN_{6}-2(N_{1}m_{3}+N_{2}m_{4}-N_{3}m_{1}-N_{4}m_{2}).
\end{eqnarray}
Applying the new ansatz (\ref{eq:new_ansatz}), we have initial conditions
based on boundary conditions,
\begin{eqnarray}
n_{1}(t,t) &=&\kappa_{1}-iM_{1}(t), \ \ \ n_{2}(t,t)=\kappa_{2}-iM_{2}(t),
\\
n_{3}(t,t) &=&-iM_{3}(t), \ \ \ \ \ \ n_{4}(t,t)=-iM_{4}(t),
\\
n_{5}(t,t,s') &=&-iM_{5}(t,s'), \ \ \ n_{6}(t,t,s')=-iM_{6}(t,s'),
\\
m_{1}(t,t) &=& -iN_{1}(t), \ \ \ \ \ \ \ m_{2}(t,t)=-iN_{2}(t),
\\
m_{3}(t,t) &=&-iN_{3}(t), \ \ \ \ \ \ \ m_{4}(t,t)=-iN_{4}(t),
\\
m_{5}(t,t,s') &=&-iN_{5}(t,s'), \ \  \ \  m_{6}(t,t,s')=-iN_{6}(t,s').
\end{eqnarray}


\section*{References}
\noindent
\begin{thebibliography}{000}



\bibitem{cqed-bec}F. Brennecke, T. Donner, S. Ritter, T. Bourdel, M. K\"{o}hl, T. Esslinger, Nature, \textbf{450}, 268 (2007).
\bibitem{cqed-qd}A. Imamoglu, H. Schmidt, G. Woods, and M. Deutsch, Phys. Rev. Lett. \textbf{79}, 1467 (1997).
\bibitem{cqed-block}A. Imamoglu, D. Awschalom, G. Burkard, D. DiVincenzo, D. Loss, M. Sherwin, and A. Small, Phys. Rev. Lett. \textbf{83}, 4204 (1999).
\bibitem{cqed-block2}C. Hamsen, K. Tolazzi, T. Wilk, G. Rempe, Phys. Rev. Lett. \textbf{118}, 133604 (2017).
\bibitem{CQED1}Y. Colombe, T. Steinmetz, G. Dubois, F. Linke, D. Hunger and J. Reichel, Nature \textbf{450}, 272 (2007).
\bibitem{CQED2}T. J. Kippenberg and K. J. Vahala, Science \textbf{321}, 1172 (2008).
\bibitem{CQED3}F. Elste, S. M. Girvin, and A. A. Clerk, Phys. Rev. Lett. \textbf{102}, 207209 (2009)
\bibitem{CQED4}J. Bernu, S. Del\'{e}glise, C. Sayrin, S. Kuhr, I. Dotsenko, M. Brune, J. M. Raimond, and S. Haroche, Phys. Rev. Lett. \textbf{101}, 180402 (2008).
\bibitem{CQED5}P. J. Leek, M. Baur, J.M. Fink, R. Bianchetti, L. Steffen, S. Filipp, and A. Wallraff, Phys. Rev. Lett. \textbf{104}, 100504 (2010).
\bibitem{CQED6}J. M. Raimond, M. Brune, and S. Haroche, Rev. Mod. Phys. \textbf{73}, 565 (2001).
\bibitem{cqed-1}C. Y. Hu, W. J. Munro, and J. G. Rarity, Phys. Rev. B, \textbf{78}, 125318 (2008).
\bibitem{cqed-2}V. Loo, L. Lanco, O. Krebs, P. Senellart, and P. Voisin, Phys. Rev. B, \textbf{83}, 033301 (2011).

%%% macroscopic qm
\bibitem{mqm1}A. D. O'Connell, M. Hofheinz, M. Ansmann, R.C. Bialczak, M. Lenander, E. Lucero, M. Neeley,
D. Sank, H. Wang, M. Weides, J. Wenner, J.M. Martinis and A. N. Cleland, Nature \textbf{464}, 697 (2010).
\bibitem{mqm2}J. Chan, T.P. Mayer Alegre, A.H. Safavi-Naeini, J.T. Hill, A. Krause, S. Gr\"{o}blacher, M. Aspelmeyer and O. Painter, Nature \textbf{478}, 89 (2011).

%%% cavity in GW
\bibitem{GWD}H. Yang, H. Miao, and Y. Chen, Phys. Rev. A \textbf{785}, 040101 (2012).
\bibitem{GWD2}H. Miao, R. Adhikari, Y. Ma, B. Pang and Y. Chen, Phys. Rev. Lett. \textbf{119}, 050801 (2017).
\bibitem{GWD3}Y. Ma, H. Miao, B. Pang, M. Evans, C. Zhao, J. Harms, R. Schnabel and Y. Chen, Nature Phys., \textbf{13}, 776 (2017).




%%% quantum measurement
\bibitem{QM1}V.B. Braginsky and F.YA. Khalili, \emph{Quantum Measurement} (Cambridge Univ. Press, 1992).


%%% quantum optics
\bibitem{QO1}M. O. Scully, M. S. Zubairy, \emph{Quantum Optics,
}(Cambridge Univ. Press, 1997).

\bibitem{QO2}M.A. Nielson and I.L. Chuang, \emph{Quantum Computation
and Quantum Information }(Cambridge Univ. Press, 2000).

%%% quantum open systems
\bibitem{QO3}H.P. Breuer, F. Petruccione, \emph{Theory of Open
Quantum Systems} (Oxford. New York, 2002).

\bibitem{QO4}C. W. Gardiner and P. Zoller, \emph{Quantum Noise}



% non-Markov
\bibitem{NM1}T. Yu and J. H. Eberly, Phys. Rev. Lett.  \textbf{93}, 140404 (2004).
\bibitem{NM2}B. L. Hu, J. P. Paz, Y. Zhang, Phys. Rev. D \textbf{45}, 2843 (1992).
\bibitem{NM3}J. P. Paz and A. J. Roncaglia, Phys. Rev. Lett. {\bf 100}, 220401 (2008).
\bibitem{NM4}W. T. Strunz and T. Yu, Phys. Rev. A \textbf{69}, 052115 (2004).
\bibitem{NM5}L. Di\'{o}si, N. Gisin and W.T. Strunz, Phys. Rev. A \textbf{58}, 1699 (1998).
\bibitem{QM2}L. Di\'{o}si, Phys. Rev. Lett. \textbf{100}, 080401 (2008).
\bibitem{MKV1} C. Anastopoulos, S. Shresta and  B. L. Hu, Quantum Inform. Process. {\bf 8}, 549 (2009).
\bibitem{NM6}J.S. Xu, C. F. Li, M. Gong, X. B. Zou, C.H. Shi, G. Chen, and G. C. Guo, Phys. Rev. Lett. {\bf 104}, 100502 (2010).
\bibitem{NM7}A. Z. Chaudhry and J. Gong, Phys. Rev. A {\bf 85}, 012315(2012).
\bibitem{chen1} Y. Chen, J. Q. You and T. Yu, Phys. Rev. A \textbf{90}, 052104 (2014).
\bibitem{chen2} Y. Chen and T. Yu, \emph{Low-Dimensional and Nanostructured Materials and Devices}, chapter 25, 609 - 633 (Springer, 2016).
\bibitem{NM8}T. Yu, Phys. Rev. A \textbf{69}, 062107 (2004).
\bibitem{chen3} Y. Chen, J. Q. You and T. Yu, arXiv:1712.06795

\end{thebibliography}
\end{document}